\documentstyle[aps,twocolumn]{revtex}

\begin{document}
\title{\begin{flushright}
\footnotesize{CECS-PHY-99/12}\\
\footnotesize{hep-th/9907109}
\end{flushright}Higher Dimensional Gravity, Propagating Torsion and AdS
Gauge Invariance}
\author{Ricardo Troncoso$^{1}$ and Jorge Zanelli$^{1,2}$}
\address{$^{1}$Centro de Estudios Cient\'{\i }ficos (CECS), Casilla 1469, Valdivia,
Chile. \\
$^{2}$Departamento de F\'{\i }sica, Universidad de Santiago de Chile,
Casilla 307, Santiago 2, Chile.}
\maketitle

\begin{abstract}
The most general theory of gravity in $d$ dimensions which leads to second
order field equations for the metric has $[(d-1)/2]$ free parameters.\ It is
shown that requiring the theory to have the maximum possible number of
degrees of freedom, fixes these parameters in terms of the gravitational and
the cosmological constants. In odd dimensions, the Lagrangian is a
Chern-Simons form for the (A)dS or Poincar\'{e} groups. In even dimensions,
the action has a Born-Infeld-like form.

Torsion may occur explicitly in the Lagrangian in the parity-odd sector and
the torsional pieces respect local (A)dS symmetry for $d=4k-1$ only. These
torsional Lagrangians are related to the Chern-Pontryagin characters for the
(A)dS group. The additional coefficients in front of these new terms in the
Lagrangian are shown to be quantized.

\end{abstract}

\section{Introduction}

The possibility that spacetime may have more than four dimensions is now a
standard assumption in high energy physics.. This underscores the need to
critically examine the minimal requirements for a consistent theory of
gravity in any dimension, including both general covariance and second order
field equations for the metric. Although many different approaches have been
followed in the generalizations to $d>4$, most models assume the simplest
generalization of General Relativity to higher dimensions, namely the
Einstein-Hilbert action. The most general action for the metric satisfying
the criteria of general covariance and second order field equations for $d>4$
is a polynomial of degree $[d/2]$ in the curvature\footnote{%
Here $[x]$ represents the integer part of $x$.}, the Lanczos-Lovelock {\bf %
(LL)} theory. The LL theory in fact refers to a family parametrized by a set
of real coefficients $\alpha _{p}$, $p=0,1,...,[d/2]$, which are not fixed
from first principles.

In this note, it is shown these parameters are fixed in terms of the
gravitational and the cosmological constants, through the requirement that
the theory possess the largest possible number of degrees of freedom. As a
consequence, the action in even dimensions has a Born-Infeld-like form,
while in odd dimensions, the Lagrangian is a Chern-Simons form for the (A)dS
or Poincar\'{e} groups. The same requirement implies that torsion may occur
explicitly in the Lagrangian only for $d=4k-1$. Each of these torsional
Lagrangians are related to the Chern-Pontryagin characters for the (A)dS
group, and the coefficients in front of them ($\beta $'s) are shown to be
quantized.

Here we adopt the first order approach, where the independent dynamical
variables are the vielbein ($e^{a}$) and the spin connection ($\omega ^{ab}$%
), which obey first order differential field equations\footnote{%
The curvature and torsion two-forms are related to $e^{a}$, and $\omega
^{ab} $ through $R^{ab}=d\omega ^{ab}+\omega _{\;c}^{a}\mbox{\tiny $\wedge$}%
\omega ^{cb}$, and $T^{a}=de^{a}+\omega _{\;b}^{a}\mbox{\tiny $\wedge$}e^{b}$%
, respectively.}. The standard second order form can be obtained if the
torsion equations are solved for the connection and eliminated in favor of
the vielbein -- this step however, cannot be taken in general because the
equations for $\omega ^{ab}$ are not invertible for dimensions higher than
four. The first order formalism has the added advantage that it can be
expressed entirely in terms of differential forms and their exterior
derivatives, without ever introducing the inverse vielbein or the Hodge $*$%
-dual.

In the next section we review extensions of gravity theory beyond the
Einstein-Hilbert action for dimensions greater than four (LL theory). In
Section III it is shown that in order to have the maximum possible number of
degrees of freedom, the $\alpha _{p}$'s must be fixed in terms of the
gravitational and the cosmological constants. In Section IV the inclusion of
torsion explicitly in the Lagrangian is explored. New torsional Lagrangians
are found, which are related to the Chern-Pontryagin characters for the
(A)dS group in $d=4k-1$ dimensions. Section V contains the discussion and
summary.

\section{Beyond the Einstein-Hilbert Action}

Assuming the spacetime geometry as given by the Einstein-Hilbert ({\bf EH})
action --with or without cosmological constant-- is the most reasonable
choice in dimensions three and four\footnote{%
In $1+1$ dimensions, in order to write an action principle it is necessary
supply the theory with an extra scalar field \cite{J-T,Brown}.}, but not
necessarily so for $d>4$. The idea that a more general theory could be
employed to describe the spacetime geometry in dimensions larger than four
--even in the absence of torsion-- was first considered some sixty years ago
by Lanczos \cite{Lanczos}. More recently, it was observed that the low
energy effective Lagrangian for gravity obtained from string theory would
have curvature-squared terms --and higher powers as well-- \cite{CHSW}.
These terms are potential sources of inconsistencies as they would in
general give rise to fourth order field equations and bring in ghosts.
However, it was soon pointed out by Zwiebach \cite{Zwiebach} and Zumino \cite
{Zumino}, that if the effective Lagrangian would contain the higher powers
of curvature in particular combinations, only second order field equations
are produced and consequently no ghosts arise. The effective Lagrangian
obtained by this argument, was precisely of the form proposed by Lanczos for 
$d=5$ and, for general $d$, by Lovelock \cite{Lovelock}.

In a more recent context, there are further clues that point in this
direction. It is expected that the low energy regime of $M$-theory should be
described by an eleven-dimensional supergravity of new type, with off-shell
local supersymmetry \cite{NG}, whose Lagrangian should contain higher powers
of curvature \cite{GV}. A family of supergravity theories that satisfy these
conditions has been proposed in \cite{trz,trz2,trz'}, and the purely
gravitational sector of those theories is an extension of ordinary EH
gravity, as described below.

\subsection{First Order Formalism}

In standard General Relativity, the metric is viewed as the fundamental
field, while the affine structure of spacetime (connection) is assumed to be
a derived concept. The link between the two structures is the vanishing of
the torsion tensor, which is often imposed as an identical, off-shell
requirement for the connection. Consequently, the spin connection has no
independent, propagating degrees of freedom and the spacetime torsion is not
dynamically determined but constrained by fiat.

A purely metric formulation would be insufficient for the description of
spinor fields because they couple the antisymmetric part of the affine
connection, and therefore they are sources for the torsion. Hence, in a
theory with fundamental spinors coupled to gravity, it is necessary that the
metric and affine properties of spacetime be treated separately. Moreover,
considering the fact that spinors provide a basis of irreducible
representations for $SO(d-1,1)$ (Lorentz$_{d}$), but not for $GL(d)$, they
must be defined on a local frame on the tangent space rather than in
relation to a coordinate system on the base manifold.

Thus, in a theory containing fermions, it is more natural to look for a
formulation of gravity in which $\omega ^{ab}$\ and $e^{a}$\ are dynamically
independent, with curvature and torsion standing on similar footing. The
first order formalism offers exactly this possibility. Indeed, when torsion
is not set equal to zero, the standard variational principles --first,
second and $1.5$ order-- are no longer necessarily equivalent. For example,
varying the action with respect to $e^{a}$ -- in the ``$1.5$ order
formalism'' \cite{pvn}, yields 
\[
\delta I=\frac{\delta I}{\delta e^{a}}\delta e^{a}+\frac{\delta I}{\delta
\omega ^{bc}}\frac{\delta \omega ^{bc}}{\delta e^{a}}\delta e^{a}, 
\]
which would reduce to the Einstein equations (second order formalism)
provided $\frac{\delta I}{\delta \omega ^{bc}}=0$. Thus, in particular, in
the presence of spinors these formalisms would be different because in the
second order case, the torsion equation is imposed as a constraint, but this
is in fact a matter of choice.

In three and four dimensions, allowing $\omega ^{ab}$ and $e^{a}$ to be
dynamically independent does not modify the standard picture in practice
because any occurrence of torsion in the action leads to torsion-free
classical solutions (in vacuum)\footnote{%
In the case of coupling to spinning matter, the torsion equations allow
expressing $\omega ^{ab}$ in terms of $e^{a}$ and the matter fields.}. In
higher dimensions, however, theories that include torsion explicitly in the
Lagrangian, cannot be related, even perturbatively, to their torsion-free
counterparts \cite{mz}.

\subsection{Higher Dimensional Gravity: Lanczos-Lovelock Theory}

The main fundamental assumptions in standard General Relativity are the
requirements of general covariance and that the field equations for the
metric be second order. Based on the same principles, the LL Lagrangian is
defined as the most general $d$-form invariant under local Lorentz
transformations, constructed with the vielbein, the spin connection, and
their exterior derivatives, without using the Hodge dual\footnote{%
Avoiding the Hogde dual guarantees that the fields $\omega ^{ab}$\ and $%
e^{a} $\ that extremize the action obey first order equations.}\cite
{Zumino,Regge,tz}.

The LL Lagrangian is a polynomial of degree $[d/2]$ in the curvature,

\begin{equation}
I_{G}=\int \sum_{p=0}^{[d/2]}\alpha _{p}L^{(p)},  \label{Lovaction}
\end{equation}
where $\alpha _{p}$ are arbitrary constants, and

\begin{equation}
L^{(p)}=\epsilon _{a_{1}\cdots a_{d}}R^{a_{1}a_{2}}\cdots
R^{a_{2p-1}a_{2p}}e^{a_{2p+1}}\cdots e^{a_{d}}.  \label{Lovlag}
\end{equation}
Here and in the sequel wedge product of forms is implicitly understood.

The first two terms in (\ref{Lovaction}) are the EH action. Although General
Relativity is contained in the LL action as a particular case, theories with
higher powers of curvature are dynamically very different from EH, whose
classical solutions are not even perturbatively related to those of
Einstein's theory. However, to lowest order in perturbation theory around a
flat, torsion-free background, all of the $L^{(p)}$'s defined in (\ref
{Lovlag}) with $p\geq 2$ are total derivatives \cite{Zumino}.

The $[(d+1)/2]$ dimensionful constants $\alpha _{p}$ in the LL action
contrast with the two constants of the EH theory ($G$ and $\Lambda $). In
the following section the $\alpha _{p}$'s are selected according to the
criterion that the integrability (or consistency) conditions for the field
equations should not impose additional algebraic constraints on the
curvature and torsion tensors. This guarantees that the fields attain the
maximum number of degrees of freedom allowed by the spacetime dimension.

\section{Consistency of the First Order Formalism}

Consider the LL action (\ref{Lovaction}), as a functional of the spin
connection and the vielbein, $I_{G}=I_{G}[\omega ^{ab},e^{a}]$. Varying with
respect to these fields, the following field equations are obtained,

\begin{eqnarray}
\delta e^{a} &\rightarrow &{\cal E}_{a}=0,  \label{E-L} \\
\delta \omega ^{ab} &\rightarrow &{\cal E}_{ab}=0,  \label{Tor}
\end{eqnarray}
where we have defined

\begin{eqnarray}
{\cal E}_{a}:=\sum_{p=0}^{[\frac{d-1}{2}]}\alpha_{p}(d-2p){\cal E}_{a}^p,
\label{e-l} \\
{\cal E}_{ab}:=\sum_{p=1}^{[\frac{d-1}{2}]}\alpha _{p}p(d-2p){\cal E}%
_{ab}^{p},  \label{tor}
\end{eqnarray}
and

\begin{equation}
{\cal E}_{a}^{p}\equiv \epsilon _{ab_{1}\cdots b_{d-1}}R^{b_{1}b_{2}}\cdots
R^{b_{2p-1}b_{2p}}e^{b_{2p+1}}\cdots e^{b_{d-1}},
\end{equation}
\begin{equation}
{\cal E}_{ab}^{p}\equiv \epsilon _{aba_{3}\cdots a_{d}}R^{a_{3}a_{4}}\cdots
R^{a_{2p-1}a_{2p}}T^{a_{2p+1}}e^{a_{2p+2}}\cdots e^{a_{d}}.
\end{equation}

The $(d-1)$-forms ${\cal E}_{a}$\ and ${\cal E}_{ab}$\ are independent
Lorentz tensors with the same number of components as the fields $e^{a}$\
and $\omega ^{ab}$, respectively. If there were algebraic relations among
these tensors, so that (\ref{E-L}) and (\ref{Tor}) would not be independent,
then the fields $\omega ^{ab}$\ and $e^{a}$\ would not be completely
determined by their field equations and initial conditions. On the other
hand, it is easy to check that by virtue of the Bianchi identities ($%
DR^{ab}=0$, $DT^{a}=R_{\text{ }b}^{a}e^{b}$), the following relations hold

\begin{equation}
D{\cal E}_{a}^{p}=(d-2p-1)e^{b}{\cal E}_{ba}^{p+1},  \label{D}
\end{equation}
for $0\leq p\leq [(d-1)/2]$, which leads to the following off-shell identity

\begin{equation}
D{\cal E}_{a}\equiv \sum_{p=1}^{[\frac{d+1}{2}]}\alpha
_{p-1}(d-2p+2)(d-2p+1)e^{b}{\cal E}_{ba}^{p},  \label{Consistency}
\end{equation}
Which by consistency with (\ref{E-L}) must also vanish. Moreover, taking the
exterior product of (\ref{tor}) with $e^{b}$ gives

\begin{equation}
e^{b}{\cal E}_{ba}\equiv \sum_{p=1}^{[\frac{d-1}{2}]}\alpha _{p}p(d-2p)e^{b}%
{\cal E}_{ba}^{p},  \label{Tors*e}
\end{equation}
which vanishes by virtue of (\ref{Tor}).

Comparing the last two identities, one can see that if the coefficients $%
\alpha _{p}$\ were generic, equations (\ref{E-L}) and (\ref{Tor}) would
imply in general additional restrictions of the form $e^{b}{\cal E}%
_{ba}^{p}=0$\ for some $p$'s. This in turn would imply that some components
of the torsion tensor must vanish, freezing out some degrees of freeom in
the theory, and at the same time, leaving other components of the curvature
and torsion tensors are left undetermined by the field equations. Thus,
different choices of $\alpha _{p}$'s correspond, in general, to theories
with different numbers of physical degrees of freedom depending on how many
additional off-shell constraints are imposed on the geometry by the last
identities.

As we show next, among all the possible choices, there is a very special one
which occurs only in odd dimensions, for which there are no additional
constraints. In even dimensions, this possibility does not exist; in fact,
equations (\ref{Consistency}) and (\ref{Tors*e}) are proportional to each
other term by term for $d=2n-1$\ but for $d=2n$, both equations have
different number of terms. We will treat each case separately.

\subsection{$d=2n-1$: Local (A)dS Chern-Simons Gravity}

In odd dimensions, equations (\ref{Consistency}) and (\ref{Tors*e}) have the
same number of terms because the last term in (\ref{Consistency}) vanishes.
Thus, if equations (\ref{Consistency}) and (\ref{Tors*e}) are to imposse no
further algebraic constraints on $R^{ab}$\ and $T^{a}$, the two series $D%
{\cal E}_{a}$ and $e^{b}{\cal E}_{ba}$ must be proportional term by term: 
\[
\gamma \alpha _{p-1}(d-2p+2)(d-2p+1)e^{b}{\cal E}_{ba}^{p}=\alpha
_{p}p(d-2p)e^{b}{\cal E}_{ba}^{p}, 
\]
which implies the following recursion relation for the $\alpha _{p}$'s

\begin{equation}
\gamma \frac{\alpha _{p}}{\alpha _{p+1}}=\frac{(p+1)(d-2p-2)}{(d-2p)(d-2p-1)}
\label{recursion}
\end{equation}
where $0\leq p\leq n-1$, and $\gamma $ is an arbitrary constant of dimension
[length$^{2}$]. The solution of this equation is

\begin{equation}
\alpha _{p}=\alpha _{0}\frac{(2n-1)(2\gamma )^{p}}{(2n-2p-1)}\left( 
\begin{array}{c}
n-1 \\ 
p
\end{array}
\right) .  \label{solution}
\end{equation}
Thus, the action contains only two fundamental constants, $\alpha _{0}$ and $%
\gamma $, related to the gravitational and the cosmological constants through%
\footnote{%
For any dimension, $l$ is a length parameter related to the cosmological
constant by $\Lambda =\pm \frac{(d-1)(d-2)}{2l^{2}}$ . In the following we
will choose the negative sign, but the analysis does not depend on it. Here,
the gravitational constant $G$ is related to $\kappa $ through $\kappa
^{-1}=2(d-2)!\Omega _{d-2}G$ (see Ref. \cite{BH-Scan}).}

\begin{eqnarray}
\alpha _{0} &=&\frac{\kappa }{dl^{d-1}},  \nonumber \\
\gamma &=&-sgn(\Lambda )\frac{l^{2}}{2}.  \label{Constants}
\end{eqnarray}

Choosing the coefficients $\alpha _{p}$\ as in (\ref{solution}), implies
that the action is invariant not only under standard local Lorentz rotations
($\delta e^{a}=\lambda _{\;b}^{a}e^{b}$\ and $\delta \omega ^{ab}=-D\lambda
^{ab}$), but also under local AdS boosts, 
\begin{eqnarray}
\delta e^{a} &=&-D\lambda ^{a}  \nonumber \\
\delta \omega ^{ab} &=&\frac{1}{l^{2}}(\lambda ^{a}e^{b}-\lambda ^{b}e^{a}).
\label{transfAdS}
\end{eqnarray}

This can be seen because the Lagrangian in (\ref{Lovaction}) with the choice
of coefficients (\ref{solution}) is the Euler-Chern-Simons form for $%
SO(d-1,2)$, that is, its exterior derivative is the Euler form in $2n$
dimensions ${\sf E}_{2n}$, 
\begin{eqnarray}
dL_{G\;2n-1}^{AdS} &=&\frac{\kappa l}{2n}\epsilon _{A_{1}\cdots A_{2n}}{\bf 
\bar{R}}^{A_{1}A_{2}}\cdots {\bf \bar{R}}^{A_{2n-1}A_{2n}}  \nonumber \\
&=&\bar{\kappa}{\sf E}_{2n},  \label{Euler}
\end{eqnarray}
where 
\begin{equation}
{\bf \bar{R}}^{AB}=\left[ 
\begin{array}{cc}
R^{ab}+\frac{1}{l^{2}}e^{a}e^{b} & T^{a}/l \\ 
-T^{b}/l & 0
\end{array}
\right] ,  \label{R}
\end{equation}
defines the Lie algebra valued AdS curvature ${\bf F}=\frac{1}{2}{\bf \bar{R}%
}^{AB}J_{AB}=d{\bf A}+{\bf A}^{2}$ in terms of the AdS connection ${\bf A=}%
\frac{1}{2}{\bf W}^{AB}J_{AB}=\frac{1}{2}\omega ^{ab}J_{ab}+e^{a}J_{ad+1}$ 
\cite{chamslett,btz}. Hence, equations (\ref{E-L}) and (\ref{Tor}) can be
cast as different components of a single AdS covariant equation 
\begin{equation}
\delta {\bf W}^{AB}\rightarrow {\cal E}_{AB}:=\epsilon _{ABA_{3}\cdots
_{d+1}}{\bf \bar{R}}^{A_{3}A_{4}}\cdots {\bf \bar{R}}^{A_{d}A_{d+1}}=0,
\label{EqCS}
\end{equation}
which transforms as a tensor under local AdS gauge transformations which
include (\ref{transfAdS}), $\delta {\bf W}^{AB}=-\nabla \lambda ^{AB}$,
where $\nabla $\ is the exterior covariant derivative in the AdS connection.
Considering this, the consistency condition $e^{b}{\cal E}_{ba}=\gamma D%
{\cal E}_{a}$\ does not produce additional constraints, because it is just a
component of the identity 
\begin{equation}
\nabla {\cal E}_{AB}\equiv 0,
\end{equation}
which is trivially satisfied by virtue of the AdS Bianchi identity, $\nabla 
{\bf \bar{R}}^{AB}=0$.

\subsection{$d=2n$: Born-Infeld-Like Gravity}

For even dimensions, equation (\ref{Consistency}) has one more term than (%
\ref{Tors*e}). Therefore, both series cannot be compared term by term and
one must follow a different route.\ It can be noted that equation (\ref{Tor}%
) is an exterior covariant derivative, 
\begin{equation}
{\cal E}_{ab}=D{\cal T}_{ab},  \label{Dtau}
\end{equation}
where,

\begin{equation}
{\cal T}_{ab}:=\frac{\delta L}{\delta R^{ab}}=\sum_{p=1}^{[\frac{d-1}{2}%
]}\alpha _{p}p{\cal T}_{ab}^{p},
\end{equation}
and 
\begin{equation}
{\cal T}_{ab}^{p}=\epsilon _{aba_{3}\cdots a_{d}}R^{a_{3}a_{4}}\cdots
R^{a_{2p-1}a_{2p}}e^{a_{2p+1}}\cdots e^{a_{d}}.  \label{Taup}
\end{equation}
Note also that ${\cal T}_{ab}^{p}$\ is related with ${\cal E}_{a}^{p}$ and $%
{\cal E}_{ab}^{p}$ through 
\begin{equation}
e^{b}{\cal T}_{ab}^{p}={\cal E}_{a}^{p-1},  \label{Tau1}
\end{equation}
\begin{equation}
D{\cal T}_{ab}^{p}=(d-2p){\cal E}_{ab}^{p},  \label{Tau2}
\end{equation}
for $1\leq p\leq [\frac{d-1}{2}]$.

Differentiating both sides of (\ref{Tau1}) and using (\ref{Tau2}), identity (%
\ref{Consistency}) can also be written for $d=2n$ as 
\begin{eqnarray}
D{\cal E}_{a} &=&T^{b}\sum_{p=1}^{n-1}2\alpha _{p-1}(n-p+1){\cal T}_{ab}^{p}
\nonumber \\
&&-\sum_{p=1}^{n-1}4\alpha _{p-1}(n-p+1)(n-p)e^{b}{\cal E}_{ba}^{p}.
\label{ConsistencyBI}
\end{eqnarray}
This equation can be compared with the second identity (\ref{Tors*e}) 
\begin{equation}
e^{b}{\cal E}_{ba}\equiv \sum_{p=1}^{n-1}2p\alpha _{p}(n-p)e^{b}{\cal E}%
_{ba}^{p}.  \label{Tors*e2n}
\end{equation}
Both (\ref{Dtau}) and (\ref{Tors*e2n}) can be zero either if $T^{a}=0$, or $%
{\cal T}_{ab}=0$. However, those are excessive conditions for the vanishing
of (\ref{ConsistencyBI}). It is sufficient to impose the weaker conditions $%
T^{a}{\cal T}_{ab}=0$, and at the same time that the second term in (\ref
{ConsistencyBI}) be proportional to the series in (\ref{Tors*e2n}). Now,
both series possess the same number of terms, and therefore the solution
which allows the maximum number of degrees of freedom is the one for which
both series are equal term by term, up to a global factor. Hence, one
obtains the following recursion relation for the $\alpha _{p}$'s: 
\begin{equation}
2\gamma (n-p+1)\alpha _{p-1}=p\alpha _{p},  \label{Rec2n}
\end{equation}
for some fixed $\gamma $. With this relation, equation (\ref{ConsistencyBI})
reads, 
\begin{equation}
D{\cal E}_{a}=\frac{1}{\gamma }(T^{b}{\cal T}_{ab}-e^{b}{\cal E}_{ab})=0,
\end{equation}
and therefore it is apparent that if $T^{a}$ is just a null vector of ${\cal %
T}_{ab}$, both consistency conditions are the same.

The solution of the recursion relation (\ref{Rec2n}) is

\begin{equation}
\alpha _{p}=\alpha _{0}(2\gamma )^{p}\left( 
\begin{array}{c}
n \\ 
p
\end{array}
\right) ,  \label{CoefsBI}
\end{equation}
with $0\leq p\leq n-1$. This formula can be extended to $p=n$ at no extra
cost, because it only amounts to adding the Euler density to the Lagrangian
with the weight $\alpha _{n}=\alpha _{0}(2\gamma )^{n}$.

The action depends only on the gravitational and the cosmological constants,
as in odd dimensions, given by (\ref{Constants}). The choice of coefficients
(\ref{CoefsBI}) implies that the Lagrangian takes the form 
\begin{equation}
L=\frac{\kappa }{2n}\epsilon _{a_{1}\cdots a_{d}}\bar{R}^{a_{1}a_{2}}\cdots 
\bar{R}^{a_{d-1}a_{d}},  \label{BI}
\end{equation}
which is the pfaffian of the $2$-form $\bar{R}^{ab}=R^{ab}+\frac{1}{l^{2}}%
e^{a}e^{b}$, and can be formally written as the Born-Infeld {\bf (BI)}-like
form \cite{jjg}, 
\begin{equation}
L=2^{n-1}(n-1)!\kappa \sqrt{\det \left( R^{ab}+\frac{1}{l^{2}}%
e^{a}e^{b}\right) }.  \label{BI'}
\end{equation}

In four dimensions (\ref{BI'}) reduces to a particular linear combination of
the Einstein-Hilbert action, the cosmological constant and the Euler
density. Although this last term does not contribute to the field equations,
it plays an important role in the definition of conserved charges for
gravitation theories in dimensions $2n\geq 4$ \cite
{charges3+1,charges2n,BH-Scan}.

The field equations (\ref{E-L}) and (\ref{Tor}), now take the form

\begin{eqnarray}
\delta e^{a} &\rightarrow &\epsilon _{ab_{1}\cdots b_{d-1}}\bar{R}%
^{b_{1}b_{2}}\cdots \bar{R}^{b_{d-3}b_{d-2}}e^{b_{d-1}}=0,  \nonumber \\
\delta \omega ^{ab} &\rightarrow &\epsilon _{aba_{3}\cdots a_{d}}\bar{R}%
^{a_{3}a_{4}}\cdots \bar{R}^{a_{d-3}a_{d-2}}T^{a_{d-1}}e^{a_{d}}=0.
\label{EqBI}
\end{eqnarray}
One could consider sectors of the phase space for which

\begin{equation}
{\cal T}_{ab}=\frac{\kappa }{2}\epsilon _{aba_{3}\cdots a_{d}}\bar{R}%
^{a_{3}a_{4}}\cdots \bar{R}^{a_{d-1}a_{d}}=0,  \label{TauBI}
\end{equation}
which solves the field equations (\ref{EqBI}) identically without requiring $%
T^{a}=0$.

The two-form $\bar{R}^{ab}$ is a piece of the AdS curvature (\ref{R}). This
fact seems to suggest that the system might be naturally described in terms
of an AdS connection (see, e.g., \cite{freund}). However, that is incorrect:
in even dimensions, the Lagrangian (\ref{BI}) is invariant under local
Lorentz transformations and {\em not} under the entire AdS group. In
contrast, as shown above, it is possible to construct gauge invariant
theories of gravity under the full AdS group in odd dimensions.

\subsection{Dynamical Behavior}

As shown above, unlike in the EH theory, the field equations of BI and CS
theories do not imply the vanishing of torsion in absence of matter. On the
contrary, assuming $T^{a}=0$ as a constraint links the transformation of the
spin connection with that of the vielbein, 
\begin{equation}
\delta \omega ^{ab}=\frac{\delta \omega ^{ab}}{\delta e^{c}}\delta e^{c}.
\end{equation}
This dynamical dependence between $\omega ^{ab}$ and $e^{a}$, spoils the
possibility of interpreting the local AdS boosts -- or local translational
invariance, in the $\Lambda \rightarrow 0$ limit -- as a gauge symmetry of
the action. Thus, the spin connection and the vielbein --the soldering
between the base manifold and the tangent space-- cannot be identified as
the compensating fields for local Lorentz rotations and AdS boosts,
respectively. Hence, gravitation can be realized as a truly gauge theory
with fiber bundle structure, where $\omega ^{ab}$ and $e^{a}$ are connection
fields only if the torsion is not fixed to zero. As shown above, this
possibility is fully realized in odd dimensions for the CS Lagrangian only.

For a generic LL theory, when torsion is set equal to zero, the number of
degrees of freedom is the same as in the EH theory, namely $\frac{d(d-3)}{2}$
\cite{tz}. These degrees of freedom correspond to the components of the
vielbein that remain after fixing the local Lorentz and diffeomorphism
invariances. On the other hand, CS theory in $d=2n-1$ without assuming
vanishing torsion has $(n-1)(2n^{2}-5n+1)$ additional degrees of freedom 
\cite{bgh}. These extra local degrees of freedom cannot be excitations
described by the vielbein, because the gauge symmetry of the theory can be
used to gauge away $\frac{d(d+3)}{2}$ of its components, just like in the
non-CS case. Hence, the extra degrees of freedom must be carried by the
contorsion tensor $k^{ab}:=\omega ^{ab}-\bar{\omega}^{ab}(e)$, where $\bar{%
\omega}^{ab}(e)$ is the solution of $T^{a}=0$.

In view of the preceding analysis, there seems to be no reason to exclude
torsion from the Lagrangian itself. In the next section we explore the
possibility of adding torsion explicitly to the action.


\section{Adding Torsion explicitly in the Lagrangian}


The generalization of the Lanczos-Lovelock theory to include torsion
explicitly is obtained assuming the Lagrangian to be the most general $d$%
-form constructed with the vielbein and the spin connection without using
the Hodge dual, and invariant under local Lorentz transformations. A
constructive algorithm to produce all possible local Lorentz invariants from 
$e^{a},R^{ab}$ and $T^{a}$ is given in Ref. \cite{mz}. As with the LL
Lagrangian, the explicit inclusion of torsional terms brings in a number of
arbitrary dimensionful coefficients $\beta _{k}$ analogous to the $\alpha
_{p}$'s.

The aim of this section is to show that in certain dimensions the $\beta $'s
can be suitably chosen so as to enlarge the local Lorentz invariance into
the AdS gauge symmetry\footnote{%
Here $SO(d)$ and $SO(d-1,1)$ will be used indistinguishably to represent the
Lorentz group in $d$ dimensions, while the $d$-dimensional (A)dS group will
be denoted by $SO(d-1,2)$, $SO(d,1)$ or $SO(d+1)$, since the analysis that
follows is insensitive to the signature.}.

If no additional structure (e.g., inverse metric, Hodge dual $(*)$, etc.) is
assumed, AdS invariant integrals can only be produced in $4k$ and $4k-1$
dimensions. This can be seen as follows: As is well-known (see, e.g., \cite
{nakahara}), in $2n$ dimensions, the only $2n$-forms invariant under $SO(N)$
constructed in terms of the $SO(N)$ curvature are the Euler density --for $%
N=2n$ only--, and the $n$-th Chern characters --for any $N$. An important
difference between these invariants is\ that the Euler form is even under
parity transformations, while the latter is odd. Parity is defined by the
sign change induced by a simultaneous inversion of one coordinate in the
tangent space and in the base manifold. Thus, for instance the Euler
density, ${\sf E}_{2n}=\epsilon _{a_{1}\cdots a_{2n}}R^{a_{1}a_{2}}\cdots
R^{a_{2n-1}a_{2n}}$, is even under parity, while the Lorentz Chern classes, $%
R_{\;b_{2}}^{a_{1}}\cdots R_{\;a_{1}}^{a_{2n}}$, and the torsional
invariants such as $e_{a}R_{\;b}^{a}T^{b}$ are parity odd.

In the previous section we discussed all possible Lagrangians of the form $%
\epsilon [R]^{p}[e]^{d-2p}$. In what follows we concentrate on the
construction of the pure gravity sector as a gauge theory which is
parity-odd. That sector is described by Lagrangians containing
Lorentz-invariant products of the fields and their exterior derivatives,
which do not contain the Lorentz Levi-Civita symbol $\epsilon _{a_{1}\cdots
a_{d}}$. This construction was sketched through briefly in the context of
supergravity in Refs. \cite{tronco,trz,trz'}.

In even dimensions the only AdS-invariant $d$-forms are, apart from the
Euler density, linear combinations of products of the type

\begin{equation}
{\sf P}_{r_{1}\cdots r_{s}}={\sf C}_{r_{1}}\cdots {\sf C}_{r_{s}},
\label{chern}
\end{equation}
with $2(r_{1}+r_{2}+\cdots +r_{s})=d$. Here 
\begin{equation}
{\sf C}_{r}=\text{Tr}({\bf F})^{r},  \label{Cr}
\end{equation}
defines the $r^{th}$ Chern character of $SO(N)$. Now, since the curvature
two-form ${\bf F}$ is in the vector representation it is antisymmetric in
the group indices. Thus, the powers $r_{j}$ in (\ref{Cr}) are necessarily
even, and therefore (\ref{chern}) vanishes unless $d$ is a multiple of four.
These results can be summarized in the following lemmas:\newline

{\bf Lemma 1:} For $d=4k$, the only parity-odd $d$-forms built from $e^{a}$, 
$R^{ab}$ and $T^{a}$, invariant under the AdS group, are the Chern
characters for $SO(d+1)$.\newline

{\bf Lemma 2:} For $d=4k+2$, there are no parity-odd $SO(d+1)$-invariant $d$%
-forms constructed from $e^{a}$, $R^{ab}$ and $T^{a}$.\newline

Since the expressions ${\sf P}_{r_{1}\cdots r_{s}}$ in (\ref{chern}) are $4k$%
-dimensional closed forms, they are at best boundary terms which do not
contribute to the classical equations (although they would assign different
phases to configurations with nontrivial torsion in the quantum theory). In
view of this, it is clear why attempts to construct purely gravitational
theories with local AdS invariance in even dimensions have proven
unsuccessful in spite of several serious efforts \cite{freund,M-M}.

The form ${\sf P}_{r_{1}\cdots r_{s}}$ can be expressed locally as the
exterior derivative of a ($4k-1$)-form, 
\begin{equation}
{\sf P}_{r_{1}\cdots r_{s}}=dL_{T\;4k-1}^{AdS}({\bf A}).  \label{ads-cs}
\end{equation}
This implies that for each collection $\{r_{1},\cdots r_{s}\}$, $%
L_{T\;4k-1}^{AdS}$ is a good Lagrangian for the AdS group ($SO(4k)$) in $%
4k-1 $ dimensions. In a given dimension, the most general Lagrangian of this
sort is a linear combination of all possible $L_{T\;4k-1}^{AdS}$'s. It can
be directly checked that these Lagrangians necessarily involve torsion
explicitly.

These results can be summarized in the following \newline

{\bf Theorem:} In odd-dimensional spacetimes, there are two families of
first-order gravitational Lagrangians $L(e,\omega )$, invariant under local
AdS transformations:\newline

{\bf a:} The Euler-Chern-Simons form $L_{G\;2n-1}^{AdS}$, in $d=2n-1$
[parity-even]. Its exterior derivative is the Euler density in $2n$
dimensions and does not involve torsion explicitly, and \newline

{\bf b:} The Pontryagin-Chern-Simons forms $L_{T\;4k-1}^{AdS}$, in $d=4k-1$
[parity-odd]. Their exterior derivatives are Chern characters in $4k$
dimensions and involve torsion explicitly. \newline

It must be stressed that locally AdS-invariant gravity theories exist only
in odd dimensions. They are $genuine$ gauge systems, whose action comes from
topological invariants in $d+1$ dimensions. These topological invariants can
be written as the trace of a homogeneous polynomial of degree $n$ in the AdS
curvature. In summary, for dimensions $4k-1$ both {\bf a-} and {\bf b-}%
families exist, and for $d=4k+1$ only the {\bf a- }family appears


\subsection{Examples for $d=2n$}


In $d=4$, the only local Lorentz-invariant $4$-forms constructed with the
recipe just described are: 
\begin{eqnarray*}
{\sf E}_{4} &=&\epsilon _{abcd}R^{ab}R^{cd} \\
L_{EH} &=&\epsilon _{abcd}R^{ab}e^{c}e^{d} \\
L_{\Lambda } &=&\epsilon _{abcd}e^{a}e^{b}e^{c}e^{d} \\
{\sf C}_{2} &=&R_{\;b}^{a}R_{\;a}^{b} \\
L_{T_{1}} &=&R^{ab}e_{a}e_{b} \\
L_{T_{2}} &=&T^{a}T_{a}.
\end{eqnarray*}

The first three terms are even under parity and the rest are odd. Of these, $%
{\sf E}_{4}$ and ${\sf C}_{2}$ are topological invariant densities (closed
forms): the Euler density and the second Chern character for $SO(4)$,
respectively. The remaining four terms define the most general pure gravity
action in four dimensions, 
\begin{equation}
I=\int_{M_{4}}\left[ \alpha L_{EH}+\beta L_{\Lambda }+\gamma L_{T_{1}}+\rho
L_{T_{2}}\right] .
\end{equation}
The first two terms can be combined with ${\sf E}_{4}$ into the Born-Infeld
form (\ref{BI'}) which is locally invariant under Lorentz, but not under
AdS. It can also be seen that if $\gamma =-\rho $, the last two terms are
combined into a topological invariant density, the Nieh-Yan form \cite{NY}.
This choice implies that the entire odd part of the action becomes a
boundary term. Furthermore, ${\sf C}_{2}$, $L_{T_{1}}$ and $L_{T_{2}}$ can
be combined into the second Chern character of the AdS group, 
\begin{equation}
R_{\;b}^{a}R_{\;a}^{b}+\frac{2}{l^{2}}(T^{a}T_{a}-R^{ab}e_{a}e_{b})=\bar{R}%
_{\;B}^{A}\bar{R}_{\;A}^{B}.  \label{NY}
\end{equation}

The form (\ref{NY}) is the only AdS invariant constructed just with $e^{a}$, 
$\omega ^{ab}$ and their exterior derivatives, and therefore there are no
locally AdS-invariant gravity theories in four dimensions.

In view of Lemmas ($1$) and ($2$), the corresponding AdS-invariant
functionals in higher dimensions can be written in terms of the AdS
curvature as linear combinations of terms like

\begin{equation}
\tilde{I}_{r_{1}\cdots r_{s}}=\int_{M}{\sf C}_{r_{1}}\cdots {\sf C}_{r_{s}},
\end{equation}
where ${\sf C}_{r}=$Tr$[(\bar{R}_{\;B}^{A})^{r}]$ is the $r$-th Chern
character for the AdS group, and $dim(M)=r_{1}+\cdots +r_{s}$ is a multiple
of four. For example, in $d=8$ the Chern characters are 
\begin{equation}
{\sf C}_{4}=\text{Tr}[(\bar{R}_{\;B}^{A})^{4}],  \label{T7a}
\end{equation}
and 
\begin{equation}
({\sf C}_{2})^{2}=\text{Tr}[(\bar{R}_{\;B}^{A})^{2}]\mbox{\tiny $\wedge$}%
\text{Tr}[(\bar{R}_{\;B}^{A})^{2}].  \label{T7b}
\end{equation}

The corresponding integrals $\tilde{I}_{4}$ and $\tilde{I}_{2,2}$ are
topological invariants that characterize the maps $SO(9)\rightarrow M_{8}$.
Furthermore, as already mentioned, $\tilde{I}_{r_{1}\cdots r_{s}}$ vanishes
if one of the $r$'s happens to be odd, which is the case in $4k+2$
dimensions. Thus, one concludes that there are no torsional AdS-invariant
gauge theories for gravity in even dimensions.

\subsection{Examples for $d=2n-1$}


The simplest example occurs in three dimensions, where there are two locally
AdS invariant Lagrangians, namely, the Einstein-Hilbert with cosmological
constant, 
\begin{equation}
L_{G\;3}^{AdS}=\frac{1}{l}\epsilon _{abc}[R^{ab}e^{c}+\frac{1}{3l^{2}}%
e^{a}e^{b}e^{c}],  \label{AdS3G}
\end{equation}
and the ``exotic'' Lagrangian \cite{witten} 
\begin{equation}
L_{T\;3}^{AdS}=L_{3}^{*}(\omega )+\frac{2}{l^{2}}e_{a}T^{a},  \label{LT3}
\end{equation}
where 
\begin{equation}
L_{3}^{*}(\omega )\equiv \omega _{\;b}^{a}d\omega _{\;a}^{b}+\frac{2}{3}%
\omega _{\;b}^{a}\omega _{\;c}^{b}\omega _{\;a}^{c}.  \label{L*3}
\end{equation}

The Lagrangians (\ref{AdS3G}, \ref{LT3}, \ref{L*3}) are the Euler, the
Pontryagin and the Lorentz Chern-Simons forms, respectively. The most
general action for gravitation in $d=3$, which is invariant under AdS is
therefore the linear combination $\alpha L_{G\;3}^{AdS}+\beta L_{T\;3}^{AdS}$%
.

For $d=4k-1$, the number of possible exotic forms grows as the number of
elements of the partition set ${\sf \pi }(k)$ of $k$, in correspondence with
the number of composite Chern invariants of the form:

\begin{equation}
{\sf P}_{\{r_{j}\}}=\prod_{r_{j}\in {\sf \pi }(k)}{\sf C}_{r_{j}}.
\label{Pr}
\end{equation}

Thus, the most general Lagrangian in $4k-1$ dimensions takes the form

\begin{equation}
\kappa L_{G\;4k-1}^{AdS}+\beta _{\{r_{j}\}}L_{T\;\{r_{j}\}\;4k-1}^{AdS},
\label{L4k-1}
\end{equation}
where $dL_{T\;\{r_{j}\}\;4k-1}^{AdS}={\sf P}_{r_{1}\cdots r_{s}}$, with $%
\sum_{j}r_{j}=4k$. These Lagrangians are not boundary terms and, unlike the
even-dimensional case, they have proper dynamics. For example, in seven
dimensions one finds 
\begin{eqnarray*}
&&L_{T\,7}^{AdS}=\beta
_{2,2}[R_{\;b}^{a}R_{\;a}^{b}+2(T^{a}T_{a}-R^{ab}e_{a}e_{b})]L_{T\;3}^{AdS}
\\
&&+\beta _{4}[L_{7}^{*}(\omega
)+2(T^{a}T_{a}+R^{ab}e_{a}e_{b})T^{a}e_{a}+4T_{a}R_{\;b}^{a}R_{%
\;c}^{b}e^{c}],
\end{eqnarray*}
where $L_{2n-1}^{*}$ is the Lorentz-CS ($2n-1$)-form, 
\begin{equation}
dL_{2n-1}^{*}(\omega )=\text{Tr}[(R_{\;b}^{a})^{n}].
\end{equation}

It should be noted that the coefficients $\kappa $ and $\beta _{\{r_{j}\}}$
are arbitrary and dimensionless. As is shown in the next section, these
coefficients must be quantized by an extension of the argument used to prove
that $\kappa $ in (\ref{Euler}) is also quantized \cite{z}. We now extend
that argument to show that the $\beta $'s in (\ref{L4k-1}) are also
quantized.


\subsection{Quantization of parameters}

Consider the action for the connection ${\bf A}$ on a $(2n-1)$-dimensional,
compact, oriented, simply connected manifold $M$ without boundary, which is
the boundary of an oriented $(2n)$-dimensional manifold $\Omega $. By
Stokes' theorem, the action for (\ref{L4k-1}) can be written as an integral
on $\Omega $,

\begin{equation}
I_{\Omega }^{AdS}[{\bf A}]=\int\nolimits_{\Omega }\left( \bar{\kappa}{\sf E}%
_{2n}+\beta _{\{r\}}{\sf P}_{\{r\}}\right) .  \label{i0}
\end{equation}
(For $d=4k+1$ the last term is absent as the ${\sf P}_{\{r\}}$'s vanish). $%
I_{\Omega }^{AdS}[{\bf A}]$ describes a topological field theory on $\Omega $
for ${\bf A}$ which should be insensitive to the change of $\Omega $ by
another orientable manifold $\Omega $' with the same boundary, i.e., $%
\partial \Omega =M=\partial \Omega $'. Thus we have

\begin{equation}
I_{\Omega }^{AdS}[{\bf A}]=I_{\Omega ^{\prime }}^{AdS}[{\bf A}%
]+\int\nolimits_{\Omega \cup \Omega ^{\prime }}\left( \bar{\kappa}{\sf E}%
_{2n}+\beta _{\{r\}}{\sf P}_{\{r\}}\right) ,  \label{i}
\end{equation}
where the orientation of $\Omega $' has been reversed. Now, $\Gamma :=\Omega
\cup \Omega ^{\prime }$ is a closed oriented manifold formed by joining $%
\Omega $ and $\Omega $' continuously along $M$. Then (\ref{i}) can be
written as

\begin{equation}
I_{\Omega }^{AdS}[{\bf A}]=I_{\Omega ^{\prime }}^{AdS}[{\bf A}]+\bar{\kappa}%
\chi [\Gamma ]+\beta _{\{r_{j}\}}{\sf p}_{\{r_{j}\}}[\Gamma ],  \label{i2}
\end{equation}
where ${\sf p}_{\{r\}}=\int\nolimits_{\Gamma }{\sf P}_{\{r_{j}\}}$.

Substituting $I_{\Omega }$ by $I_{\Omega ^{\prime }}$ would have no effect
in the path integral for the quantum theory provided the difference $\Delta
[\Gamma ]=I_{\Omega }-I_{\Omega ^{\prime }}$ is an integer multiple of
Planck's constant $h$ which, in addition, cannot change under continuous
deformations of the fields. Thus, we have 
\begin{eqnarray}
\Delta [\Gamma ] &=&\bar{\kappa}\chi [\Gamma ]+\beta _{\{r_{j}\}}{\sf p}%
_{\{r_{j}\}}[\Gamma ] \\
&=&mh.  \nonumber  \label{q}
\end{eqnarray}
Now, since the Euler and the Pontryagin numbers $\chi [\Gamma ]$ and ${\sf p}%
_{\{r_{j}\}}[\Gamma ]$ are integers, the coefficients $\bar{\kappa}$ and $%
\beta _{\{r_{j}\}}$ are necessarily quantized.

The preceding argument is rigorously valid for a manifold with Euclidean
signature. If $M$ is locally Minkowskian one can apply the same reasoning to
the analytic continuation of the path integral in which the base manifold $M$
and its tangent bundle $T(M)$ are simultaneously Wick-rotated. This has the
effect that the group of rotations on $T(M)$ may have nontrivial homotopy
group, $\pi _{2n-1}[SO(2n)]$ so that the Chern characters can be nonzero.


\section{Discussion and Summary}

\subsubsection{Exact Solutions}

It is apparent from the field equations for CS and BI theories (\ref{EqCS},%
\ref{EqBI}), that any locally AdS spacetime is a solution for them. Apart
from anti-de Sitter space itself, some interesting spacetimes of negative
constant curvature are topological black holes of Refs. \cite{ABHP+BGM}.
Furthermore, for each dimension, there is a unique static, spherically
symmetric, asymptotically AdS black hole solution \cite{btz}, as well as
their topological extensions \cite{Cai-Soh}. Similarly,
Friedman-Robertson-Walker cosmologies have also been found\cite{IL+IKL}.

Torsional AdS-invariant terms can be coupled only for CS theory in $4k-1$
dimensions. The contributions of these terms to the field equations, vanish
identically for the solutions just mentioned --but not for all solutions.

\subsubsection{Local AdS Symmetry and Geometric Equivalence}

It may be stressed that the field equations in the CS case, with or without
torsional terms, are manifestly locally AdS-covariant, which gives rise to a
paradoxical situation: Under the action of an AdS transformation which is
not contained in the Lorentz subgroup (\ref{transfAdS}), the curvature and
torsion tensors transform as: 
\begin{eqnarray}
\delta \bar{R}^{ab} &=&\frac{1}{l^{2}}(\lambda ^{a}T^{b}-\lambda ^{b}T^{a}),
\nonumber \\
\delta T^{a} &=&-\bar{R}_{\;b}^{a}\lambda ^{b}.
\end{eqnarray}

Thus, in general, a solution with non-vanishing AdS curvature and zero
torsion can be mapped into another one with torsion. These solutions are not
diffeomorphically equivalent to each other. In fact, the metric transforms
under (\ref{transfAdS}) as

\begin{equation}
\delta g_{\mu \nu }=\delta _{\xi }g_{\mu \nu }-\xi ^{\lambda }e_{a(\nu
}T_{\mu )\lambda }^{a},  \label{delta}
\end{equation}
where $\delta _{\xi }$ stands for a diffeomorphism whose parameter satisfies 
$\lambda ^{a}=e_{\mu }^{a}\xi ^{\mu }$. This implies that in the presence of
torsion, the new metric is in general not diffeomorphic to the old one.
Furthermore, even if there is no torsion to begin with, by virtue of (\ref
{delta}) the new metric will eventually be diffeomorphically inequivalent to
higher order in $\lambda ^{a}$.

At first glance, it would seem that these two solutions should be physically
inequivalent; in fact, these solutions have different geodesic structure.
The apparent paradox stems from the fact that the geodesic equation is
Lorentz covariant, but not AdS covariant. The crucial point, is what one
means by ``physically equivalent''. The situation is analogous to the
transformation of the electromagnetic field under a Lorentz boost: ${\bf E}$
and ${\bf B}$ fields transform, and even if one start with a purely magnetic
(electric) configuration, to second order in $(v/c)$ one finds both.


\subsubsection{String-induced Gravity}

In the context of string theory, as shown in \cite{Zwiebach,Zumino}, the LL
action is the only ghost free candidate for a low energy gravitational
effective theory. This is a consequence of the fact that this are the only
theories which gives rise second order field equations for the metric. On
the other hand, it has been argued that the on-shell $S$-matrix is unchanged
under metric redefinitions $g_{\mu \nu }$\ $\rightarrow $\ $g_{\mu \nu
}+\alpha R_{\mu \nu }+\beta R$ \cite{T+MT+GS}, which changes completely the
polynomial structure of the effective Lagrangian. This in general brings in
ghosts, and worst yet, modifies the dynamical structure of the classical
theory: the new field equations are fourth order, and they no longer pose a
well defined Cauchy problem and spoils the causality features.

On the other hand, similar field redefinitions of the form $A_{\mu
}\rightarrow A_{\mu }(A)$\ are not acceptable in gauge theories since this
would severely damage gauge invariance, spoiling essential features of the
quantum theory. In the absence of a quantum theory of gravity, there seems
to be no way to fix the form of the action completely unless the theory
could be formulated as a gauge theory with fiber bundle structure. This is
precisely the case for the CS gravity theories, whose simplest example
occurs in $2+1$\ dimensions. Although this gauge invariance of $2+1$\
gravity is not always emphasized, it lies at the heart of the proof of
integrability of the theory \cite{witten}. Higher dimensional CS actions
have no dimensionful constants when written in terms of the AdS connection,
so that the fields have canonical dimension $1$ and the action describes a
bona fide AdS gauge system. The corresponding quantum theory as well as its
local supersymmetric extensions would be renormalizable by power counting
and possibly finite \cite{z}.

\subsubsection{Supergravity}

The analysis of stability and positivity of the energy for these theories is
a nontrivial problem. However some insights can be gained from the
supersymmetric extension of the CS theory, for which, the expectation values
of different charges should be related by Bogomolny'i bounds. The
supersymmetric extension of gravity theories described here for $d=4k-1$ was
discussed in \cite{trz}, and in general for $d=2n-1$, in \cite{trz2,trz'}.
The resulting supergravity theories are locally invariant under the
supersymmetric extensions of AdS, so that supersymmetry is realized in the
fiber rather than on the base manifold\footnote{%
An exceptionally simple case occurs when the coefficients $\alpha _{p}$ in
the theory are chosen so that the bosonic system is locally Poincar\'{e}
invariant. The supersymmetric extension was constructed in Ref. \cite{btrz}.}%
. A key point in that construction is that supersymmetry requires the
inclusion of torsional terms from the start, which justifies considering
such terms in the purely bosonic theory. The possible connection between the
new eleven dimensional supergravity and M-Theory is an open problem, as can
be readily inferred from a cursory review of Refs.\cite
{CTZ,Horava,Paniak,Hewson,Nishino,Max,Smolin}.

The first example of a supergravity action containing the LL-action was
worked out by Chamseddine in five dimensions \cite{chamslett}. This
construction, however cannot be generalized to arbitrary higher odd
dimensions unless torsional terms are introduced in the gravitational sector.

One can show that, around an appropriate background, the conserved charges
satisfy a central extension of the super AdS algebra, which leads to a
Bogomolny'i bound on the bosonic charges. As usual, solutions with Killing
spinors saturate the bound\footnote{%
The five-dimensional case was analyzed partially in \cite{CTZ}}.

In $d=2n-1$, one should expect the existence of a new kind of $2n$%
-dimensional theory at the boundary. That theory should be constructed on
the generalization of the centrally extended gauge algebra, which can be
viewed as the superconformal algebra at the boundary\footnote{%
We thank Marc Henneaux for fruitful discussions about this point.}. These
theories should be a rich arena to test the recently conjectured AdS/CFT
correspondence \cite{Maldacena}.

\subsubsection{Summary}

In sum, it was shown that requiring LL theories to have the maximum possible
number of degrees of freedom, fixes their $[(d-1)/2]$ free parameters in
terms of the gravitational and the cosmological constants. Following this
criterion, the selected theories fall into two families. In even dimensions,
torsion can be assumed to be a null vector of ${\cal T}_{ab}$ defined in (%
\ref{TauBI}), which is in general much weaker than imposing $T^{a}=0$ by
fiat, so that the resulting theory has a Born-Infeld form. In odd
dimensions, torsion needs not be constrained at all in the theory, and the
action can be written as a CS form. In that case, the vielbein and the spin
connection can be viewed as different components of an (A)dS or Poincar\'{e}
connection, so that its local symmetry is enlarged from Lorentz$_{d}$ to
(A)dS$_{d}$ (or Poincar\'{e} when $\Lambda =0$)\cite{Comment}.

The existence of propagating degrees of freedom --associated with the
spacetime contorsion $k^{ab}$--, makes it natural to look at the parity -odd
sector of the theory, which implies to consider torsional terms explicitly
in the Lagrangian. The explicit inclusion of torsional terms brings in a
number of new arbitrary dimensionful coefficients $\beta _{\{r\}}$ analogous
to the $\alpha _{p}$'s in the LL Lagrangian. It was shown that $\beta $'s
can be suitably chosen so as to enlarge the local Lorentz invariance into
the AdS gauge symmetry. In that case, torsion may occur explicitly in the
Lagrangian in the parity-odd sector and for $d=4k-1$ only. These torsional
Lagrangians are related to the Chern-Pontryagin characters for the (A)dS
group. Thus, for $d=4k-1$, the most general theory which allow the existence
of independent propagating degrees of freedom for the contorsion, has a new
set of parameters --the $\beta _{\{r\}}$'s--, which are shown to be
quantized.\newline

{\bf Acknowledgments }\newline

The authors are grateful to R. Aros, M. Contreras, J. Cris\'{o}stomo, J.
Gamboa, C. Mart\'{\i }nez, F. M\'{e}ndez, R. Olea, M. Plyushchay, J.
Saavedra, A. Sen, C. Teitelboim and B. Zwiebach for many enlightening
discussions. We are specially thankful to M. Henneaux for many helpful
comments and for his kind hospitality in Brussels. J.Z. also wishes to thank
The Abdus Salam ICTP and the organizers of the ``Extended Workshop in String
Theory'' for hospitaliy while writing up this work. This work was supported
in part through grants 1990189, 1980788 from FONDECYT.\ Institutional
support from a group of Chilean companies (CODELCO, Empresas CMPC and
Telef\'{o}nica del Sur) is also recognized. CECS is a Millenium Science
Institute.


\begin{references}
\bibitem{Comment}  BI and CS theories belong to a wider family of gravity
theories described by the choice of coefficients 
\[
\alpha _{p}=\frac{\alpha _{0}d(2\gamma )^{p}}{(d-2p)}\left( 
\begin{array}{c}
k \\ 
p
\end{array}
\right) , 
\]
which are determined requiring the existence of a unique maximally symmetric
background solution. The Einstein-Hilbert action is recovered for $k=1$, and
BI and CS Lagrangians correspond to $k=n-1$, and can be shown to possess
well behaved black hole solutions for all $k$ \cite{BH-Scan}.
\end{references}
\end{document}